\documentclass[letterpaper margin=0.5in]{article}
\usepackage{amsmath,amsfonts,amsthm,pictexwd}
\usepackage{enumerate,here}
\usepackage{graphicx}
\usepackage{fancyhdr}
\usepackage[english]{babel}
\addto{\captionsenglish}{}
\usepackage{blindtext}
\usepackage{multicol}
\usepackage{arydshln}
\usepackage{pdflscape}
\usepackage{color}
\usepackage{float}
\usepackage[numbers]{natbib}
\usepackage{url}

\def\l{{\lambda}}

\newcommand{\wegermee}[1]{}
\newcommand{\hallo}[1]{}

\begin{document}
\title{Elementary Statistics on Trial\\
(the case of Lucia de Berk)}
\author{Richard D. Gill, Piet Groeneboom and Peter de Jong}
\date{}
\maketitle

\begin{multicols}{3}
The trial of the Dutch nurse Lucia de Berk, suspected of several murders and attempts of murder was a very high profile case in the Netherlands. The initial suspicion rested mainly on quasi-statistical considerations, which produced (based partly on incorrect calculations) extremely small probabilities. Since the outcomes proved controversial, the court claimed to have dropped the statistical calculations from the verdict. But the verdict still rested on intuitive notions as ``very improbable''. So statistics were at center stage.

In the conviction of Lucia de Berk an important role was played by a simple (so-called) hypergeometric
model, used by the law psychologist (H.\ Elffers) consulted as statistician by the court, which produced very small probabilities of
occurrences of certain numbers of incidents.

In this article we want to draw attention to the fact that, if we take
into account the variation among nurses in incidents they experience during their shifts, these probabilities become considerably larger. This points to the danger of using an
oversimplified  discrete probability model in these circumstances.

The outcomes of applying our alternative model to this case are in striking contrast with those of the first calculations which led to the initial suspicions and were instrumental in determining the atmosphere surrounding the trial and subsequent hysteria. 
The main result is that under the assumption of heterogeneity, the probability of experiencing a number of incidents ($14$) that led to Lucia's conviction is about $0.0206161$ or {\it one in} 49 if the calculations are based on the same data as used by the law psychologist of the court. In his calculation, however, this probability was equal to one in 342 million.

\end{multicols}

\begin{figure}[!ht]
\begin{center}
\includegraphics[scale=0.4]{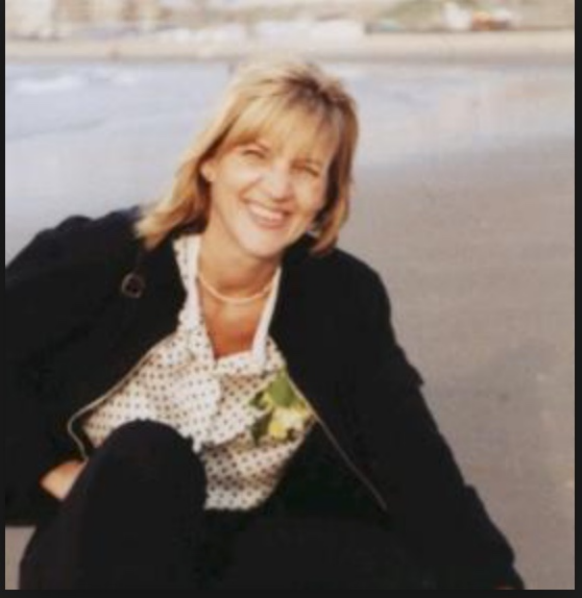}
\end{center}
\label{fig:Chernoff}
\caption{Lucia de Berk before her imprisonment}
\end{figure}

\begin{multicols}{3}
\section*{The data}
\vspace{0.2cm}
We use data from the unpublished reports of Elffers \cite{Elffers:2002_29_mei_dutch} and \cite{Elffers:2002_8_mei_dutch}. But before going into this, we want to make some general remarks on the data collection.

One of the key features of the data was the flawed data collection. Here different disciplines came into conflict: criminal investigation and scientific data gathering are very different. Their objectives, methods and results are not compatible. Criminal investigation is started when there is (suspicion of) a crime, hence one is looking for or  hunting down a suspect.
If there is need for meaningful statistics another methodology is needed, guaranteeing clear definitions and uniformity of the data collection.
In the case of Lucia de Berk this clash of cultures proved disastrous. Incidents outside shifts of Lucia were discarded and some initially reported incidents were later relabeled without clear reasons.  Extra shifts without incidents and incidents outside shifts of Lucia were subsequently brought to light. Moreover, the data collection rested for a large part on memory.

Clearly, the context of a criminal investigation produces a specific mindset: on the one hand the witnesses know what is looked for (and some of them may already be convinced of the guilt of the suspect), on the other hand fear of implicating one's self and friends can considerably distort memory.
The data on shifts and incidents for the period which was singled out in Elffers' reports are given in the following table (our Table 1 corrects an error in \cite{Meester_ea:2007}, where the number of shift in the ward RCH-41 was erroneously given by 336 instead of 366 in their table on top of p. 235).
\end{multicols}

\begin{table}[H]
	\caption{Data on shifts and incidents}
	\vspace{0.2cm}
	\centering
	\begin{tabular}{l|ccc|c}
		\hline
		\hline\
		Hospital name (and ward number) & JCH & RCH-41 & RCH-42 &Total\\
		\hline
		Total number of shifts &1029    &366   & 339 & 1734\\
		Lucia's number of shifts& 	142    &1   & 58 &201\\
		Total number of incidents &		8    &5   & 14  &27 \\
		Number of incidents during Lucia's shifts &8  &1	&5  &14\\
		\hline
		\hline
	\end{tabular}
	\label{table:table_elffers}
\end{table}

JCH and RCH denote the ``Juliana Children's Hospital" and ``Red Cross Hospital", respectively, and 41 and 42 were different ward numbers of the Red Cross hospital.

\begin{multicols}{3}
\section*{Elffers' method}
\label{method_Elffers}
We first discuss the analysis of the law psychologist H. Elffers, the statistician consulted by the court. This analysis was based on Table \ref{table:table_elffers}. As was noticed later, Lucia de Berk had actually done three shifts in RCH-41 instead of just one, but we will argue from the data used by H.\ Elffers. As explained in \cite{Meester_ea:2007}, Elffers argued by {\it conditioning} on part of the data and used two fundamental assumptions:
\begin{enumerate}
\item There is a fixed probability $p$ for the occurrence of an incident during a shift (for example, $p$ does not depend on whether the shift is a day shift or a night shift or on the nurse involved, etc.),
\item Incidents occur independently of one another.
\end{enumerate}
On the basis of these assumptions, one can compute the probability that $L$ incidents occur during Lucia's shifts, given the total number $I$ of incidents and the total number $N$ of shifts considered in the period of study. This is a {\it hypergeometric probability} given by
\begin{equation}
\label{eq:hypergeometric_prob}
\frac{\binom{S}{L}\binom{N-S}{I-L}}{\binom{N}{I}}
\end{equation}
where $S$ is the number of shifts of Lucia and $I$ is the total number of incidents, and where $\binom{S}{L}$, etc.\ denote binomial coefficients.
If we just take all the data of Table 1 together, we have a total number of $N=1734$ shifts, Lucia had $S=201$ shifts, there was a total number $I=27$ of incidents, and $L=14$ incidents during shifts of Lucia. If we evaluate (\ref{eq:hypergeometric_prob}) with these values for $N, S, I$ and $L$, we get the very small probability of about one in 4.2 million.
\end{multicols}
\begin{figure}[!ht]
\begin{center}
\includegraphics[scale=1]{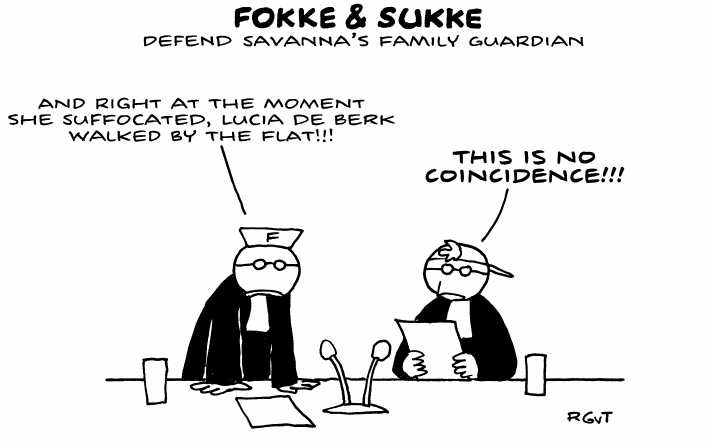}
\end{center}
\label{fig:fokke_sukke}
\caption{A Fokke \& Sukke cartoon from 10-30-2007 in the Dutch newspaper NRC Next. The text was kindly translated into English for us by the creators of the cartoon: Reid, Geleijnse and Van Tol. Lucia de Berk was still in prison at that time. The canary and the duck are defending a family guardian, accused of being responsible for the death of the girl Savanna, who died by suffocation. The accused woman was in fact acquitted (by another defense!). What counselor Sukke is saying corresponds to what the law psychologist H.\ Elffers told the court: ``Honored court, this is no coincidence. The rest is up to you.''.}
\end{figure}

\begin{multicols}{3}
If we want to compute the probability ($p$-value) that a nurse is present with $14$ {\it or more} incidents in Elffers' method of testing a null hypothesis of no systematic effects on these combined data (but he actually did not test it in this way on the combined data, see below), we have to sum the probabilities for $L=14, 15,\dots,27$, and then we get the probability of about one in $3.8$ million. This is a very small probability, although still about $100$ times larger than the probability Elffers arrived at as described below. For the model we introduce in the next section, however, we get, using the same data, a probability of {\it one in} 49.

However, Elffers proceeded somewhat differently, not combining the data of the different hospitals. The details of what he actually did are described in \cite{Meester_ea:2007}. The most important mistake he made in his calculation was to take the three hospitals separately, and multiplying the probabilities he got for these separately. This has the absurd consequence that a nurse working in several different hospitals gets a higher chance of being accused of inexplicably being present at incidences than a nurse working in just one hospital. In this way he arrived at his estimate that the probability that Lucia de Berk was present at the given numbers of incidents at the Juliana Children's hospital and the Red Cross hospital was equal to one in 342 million. We refer the interested reader to \cite{Meester_ea:2007} and to Chapter 7  ``Math error number 7: the incredible coincidence'' of the book \cite{schneps:13}.

\subsection*{Post-hoc testing}
A reviewer of this paper has asked us to comment on the issue of the danger of post-hoc testing: testing a hypothesis using the same data which suggested that hypothesis. Elffers actually tried to take account of this problem in the following way. He started from the assumption that the number of incidents in the data from JCH was much larger than expected, and that the purpose of his analysis was to discover whether there was an association with any of the nurses who worked on the ward. He multiplied his p-value for the association with Lucia's shifts by 27, the number of nurses in that period who worked on the same ward. By the time he came to look at the data from RCH, Lucia was a prime suspect and he judged that no further Bonferroni type correction was required. Finally, he proposed to take a very small probability for the significance level of his test.

In fact, his starting assumption was false: in the previous year there had been no incidents in the ward, but the year before that, an even larger number. The hospital director had not revealed the information from two years ago to the investigators since the ward previously had had a different name (he had changed it).

One could try to use a Bayesian approach to deal with the post-hoc problem. There would be good arguments for a rather low prior probability of an arbitrary nurse being a serial killer. The difficult task for the Bayesian would be determining a reasonable model for number of incidents if Lucia is a murderer, since one has to take into account that some proportion of the incidents are not murders at all. Heterogeneity would also remain an issue for a Bayesian analysis. Explaining the methodology in a court of law could well be the biggest barrier.

\end{multicols}

\begin{multicols}{3}
\section*{Alternative model}
We can model the incidents that a nurse experiences by a so-called  Poisson process, with a nurse-dependent intensity $A$, where we use $A$ for  ``accident proneness''. A Poisson process is used to model incoming phone calls during non-busy hours, fires in a big city, etc. Since we believe the incidents to be rare, a Poisson process is an obvious choice for modeling the incidents that a nurse experiences.

This approach models two separate phenomena. Firstly, the intensity of nurses seeing or reporting incidents  is modeled by introducing the random variable $A$. We assume that $A$ has an exponential distribution, but other choices are also possible.

Note that we move away here from a simple discrete model, as used by Elffers, but use instead  a {\it continuous} distribution for the ``accident proneness'' $A$ of the nurse. Statistical models with continuously varying random variables are perhaps more difficult to explain to the judges, but are often much more realistic, which should be the only important consideration here.

Secondly, the number of incidents happening to a nurse on duty depends on $A$ and the time interval she is working, and follows (conditionally on $A$) a Poisson distribution. The time interval is measured by the number of shifts the nurse has had.

Assuming that $A$ is exponentially distributed  implies, among other things, that it can easily happen that one nurse has twice the incident rate of another nurse. The probability of this event is $2/3$; in fact the probability of an incidence rate of a factor $k$ times that of another nurse is $2/(k+1)$.

The statistical problem boils down to the estimation of the parameter, characterizing the mixture of Poisson processes for the different nurses. 
 Combining the Juliana Children's Hospital and the two wards of the Red Cross Hospital, Lucia had $201$ shifts and $14$ incidents.
 
A major flaw in the investigation is that the data collection is irreproducible and lacks rigorous methods and definitions. It crucially depended on the memory of people who knew what was sought after. But we will argue from the data in Table \ref{table:table_elffers} above, which also was used in the computations of Elffers.

We'll take the overall probability of an incident per shift to be the ratio  of total number of incidents to total number of shifts, $\mu=27/1734$. If we take a shift to be our unit time interval, then this would be a so-called {\it moment estimate} of the mean intensity of incidents.

\begin{figure}[!ht]
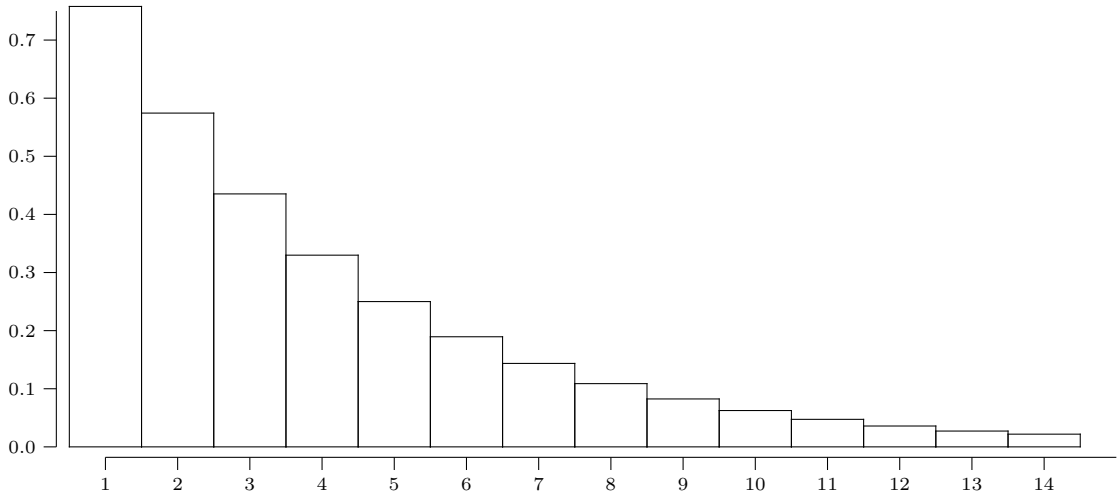

\begin{center}
\beginpicture
\scriptsize
\setcoordinatesystem units <\textwidth,\textheight>
\put {
\beginpicture
    \setcoordinatesystem units <0.06\textwidth,0.4\textheight>
    \setplotarea x from 1 to 15, y from 0 to 0.75
    \axis bottom shiftedto y=-0.018
    ticks numbered from 1 to 14 by 1
    /
\axis left shiftedto x=0.32
ticks numbered from 0 to 0.7 by 0.1
    /
\setsolid
\sethistograms
\plot
0.5 0
1.5 0.757855
2.5 0.574344 
3.5 0.43527 
4.5 0.329871
5.5 0.249995 
6.5 0.18946
7.5 0.143583
8.5 0.108815
9.5 0.0824661 
10.5 0.0624973
11.5 0.0473639
12.5 0.035895
13.5 0.0272032
14.5 0.0218641
/
\endpicture} [ t] at 0.5 0
\endpicture
\caption{Probabilities (in the Poisson model) that the number of incidents in $201$ shifts for one nurse is
at least 1,2,3,$\dots$, if $\mu=27/1704$. The probabilities are given by the heights of the columns above
$1,2,3,\dots$, respectively.}
\label{fig:Probabilities}
\end{center}
\end{figure}

This means, that, conditionally on the time interval $T=201$, the number of incidents follows a mixture of Poisson random variables with parameter $201 A$, where the intensity $A$ has an exponential distribution with first moment $\mu$. Thus on average, an innocent Lucia would experience $201\cdot\mu=201\cdot27/1734~\approx ~3.12976$ incidents. A picture of the probabilities that the number of incidents is greater or equal to $k=1,2,\dots$ is shown in Figure \ref{fig:Probabilities}, which is based on the calculations given at the end of this section.

Heterogeneity of any kind increases the variation in the number of incidents experienced by a randomly chosen nurse over a given period of time (given number of shifts). From the well-known relations  for conditional expectation ($E$) and variances ($\mathrm{var}$)
\begin{align*}
\mathrm E(X)=\mathrm E(\mathrm E(X|Y)),
\end{align*}
\begin{align*}
\mathrm {var}(X)&=
\mathrm E(\mathrm {var}(X|Y))\\
&\qquad+\mathrm {var}(\mathrm E(X|Y)),
\end{align*}
it follows that whereas for a Poisson distributed random variable variance and mean are equal, for a mixture of Poisson's (with different conditional means), the variance is larger than the mean. So if some nurses experience more or less incidents than other, in all cases the end-result is \emph{overdispersion} caused by \emph{heterogeneity}.

Applied to the current model which is geometric with parameter $(1 + t\mu)^{-1}$ (see the computation at the end of this section):
\begin{align*}
\mathrm {var}(N)& = (1 + t\mu)^2 - (1 + t\mu) \\
& =~ t\mu + (t\mu)^2, 
\end{align*}
where the latter term neatly splits over the expected variance of the Poisson process plus the variance of the conditional parameter of the Poisson process which we assumed to be exponential.

The fact that a modest amount of heterogeneity turns an almost impossible occurrence into something merely mildly unusual, is strong support for further empirical research whether and if so in what forms heterogeneity plays a role in healthcare. It can have major implications in different areas, such as medical research (representing an extra source of variation) and training of medical staff. 

\end{multicols}

\subsection*{Computation of the probabilities in the mixed Poisson model}
If $N$ is a Poisson random variable with parameter $\l$, the probability that the number of incidents is bigger than $k$, $k=1,2,\dots$,  is given by an integral, namely
\begin{equation}
\label{Poisson_exceedance}
\frac1{(k-1)!}\int_0^{\lambda}e^{-x}x^{k-1}\,dx,
\end{equation}
see, e.g., \cite{feller1:68}, Exercise 46, p.\ 173. This means that if we assume that the ``accident proneness" of the nurses has an exponential distribution with expectation $\mu$ (in our case estimated by $27/1734$) and the parameter of the Poisson distribution for the nurse is given by $ta$, where $t$ is the time interval (in our case $t=201$) and $a$ the accident proneness, we have to integrate (\ref{Poisson_exceedance}) with respect to the density of the exponential distribution with expectation $\mu$, taking $\l=ta$ So we get for the probability that a nurse experiences $k$ or more incidents:
\begin{align*}
&\int_0^{\infty}{\mathbb P}\left\{I\ge k|A=a,\,T=t\right\}\frac{e^{-a/\mu}}{\mu}\,da
=\int_0^{\infty}\left\{\frac1{(k-1)!}\int_0^{ta}e^{-x}x^{k-1}\,dx\right\}\frac{e^{-a/\mu}}{\mu}\,da\\
&= \left(\frac{t\mu}{1 + t\mu}\right)^k.
\end{align*}

\begin{multicols}{3}
This is the geometric distribution with parameter $1/(1 + t\mu)$. 
With $k = 14$ and $t\mu = 3.12976$ this yields $0.0206161$ or about \emph{one in} 49.

An early version of this paper used a revision of Elffers' data-set proposed by Professor Ton Derksen, philosopher of science, who together with his sister, medical doctor Metta de Noo, was the first to actively contest the court's reasoning in the case of Lucia de Berk.\\
\indent
Our model then led us to a right tail probability of {\it one in nine}. We later noticed that Derksen had also removed all incidents which the court finally decided not to count as provenly caused by Lucia; he used the legal argument that Elffers had previously been instructed by the judges to do the same for the data from the Juliana Children's Hospital. This does not make any statistical sense.\\ 
\indent
Going back to original medical records, Derksen and de Noo also found inconsistencies in the classification and timing of several incidents, which underlines the unreliability of the data. Correcting the data for apparent errors would also improve the results from the defence point of view. \\
\indent
We decided in the present paper to stick with Elffers' numbers in order to focus on our main point concerning the impact of heterogeneity.

\section*{Extended discussion of heterogeneity}\label{hetero}
We showed that a modest amount of heterogeneity leads to very different orders of magnitude in the outcomes of crucial calculations. Here we address some of the underlying mechanisms which may lead to the postulated heterogeneity. 

Clearly, the data in this case show heterogeneity. The data stem from two hospitals with very different patients, young children in the JCH and elder adult patients in the RCH. The data come from three wards and the rates of incidents per shift vary considerably for each ward.

We describe two general mechanisms causing heterogeneity. The first one concerns properties of subjects directly related to the intensity of the rate of incidents. The other mechanism is more indirect and results from ``spurious correlations'', in which properties not related to the underlying intensity influence the measurement via unexpected dependencies and systematic variations in variables assumed to be independent and uniform. 

Related to this is another aspect of the data: the degree to which a specific model or null-hypothesis is susceptible to small variations in the data. We will show this to be the case in the original calculations. Although our example is tuned to this very specific case, it refers to a much more general \emph{caveat}. It should be established how stable certain models are under small perturbations of the data. 

\begin{figure}[!ht]
\begin{center}
\beginpicture
\scriptsize
\setcoordinatesystem units <\textwidth,\textheight>
\put {
\beginpicture
    \setcoordinatesystem units <0.1\textwidth,0.00004\textheight>
    \setplotarea x from 0 to 8, y from 0 to 10000.000
    \axis bottom shiftedto y=-0.018
    ticks numbered from 0 to 8 by 1
    /
\axis left shiftedto x=-0.32
ticks numbered from 0 to 10000 by 1000
    /
\setsolid
\sethistograms
\plot
0 0
1 9043.864
2 1137.586
3 257.5384
4 79.49707
5 29.98891
6 13.05167
7 6.328683
8 3.341080
9 1.889573
10 0.043864
/
\endpicture} [ t] at 0.5 0
\endpicture
\caption{Robustness of the hypergeometric model, with respect to small variations in the reported suspicious incidents outside Lucia's shifts. We show the odds (divided by 1000) of 8 incidents happening inside Lucia's shifts, with k incidents outside her shift, $k=0,1, \cdots, 8$.}
\label{fig:odds}
\end{center}
\end{figure}

\subsection*{Are nurses interchangeable?}
\label{subsection:interchangeability}
According to medical specialists we have spoken to, nurses are completely interchangeable with respect to the occurrence of medical emergencies among their patients\wegermee{: nurses merely carry out the instructions given to them by the medical staff, and they do this according to standard practices of proper care, so it can make no difference at all to replace one nurse by another}. However, according to nursing staff we have consulted, this is not the case at all. Different nurses have different styles and different personalities, and this can and does have a medical impact on the state of their patients. Especially regarding care of the dying, it is folk knowledge that terminally ill persons tend to die preferentially on the shifts of those nurses with whom they feel more comfortable. \wegermee{(This might apply to the Red Cross Hospital, where Lucia worked on two adjacent wards for terminally ill aged patients).} As far as we know there has been no statistical research on this phenomenon. 

There is another obvious way in which the intensity of incidents depends on characteristics that vary over the population. Any event that can turn out to be an ``incident'' starts with the call of a doctor. And in all cases it is the nurse who decides to call a doctor. This decision is influenced by professional and personal attitude, past experience and personality traits like self-confidence. It seems obvious to us that these characteristics vary greatly in any population. Hence we assume that the intensity of experiencing incidents varies accordingly.

\subsection*{Inadequacy of the hypergeometric distribution as a model and spurious correlations}
\label{subsection:hypergeometric}
	\wegermee{Above we have mentioned two ways in which a particular nurse could have a causal but ``innocent'' influence on the occurence of an incident (since if she is replaced by another nurse it happens later or not at all). The underlying cause is unmeasured, and indeed perhaps unmeasurable, heterogeneity between nurses.}
The model underlying the null-hypothesis (which led to the hypergeometric distribution) depends on two assumptions: Both the incidents and the nurses are assigned to shifts uniformly and independently of each other. 
 
Above we have established two ways in which characteristics of individual subjects may lead to variation in the intensity of experiencing an incident. This variation is in contrast with one of the assumptions underlying the hypergeometric distribution: uniformity. 

Next we discuss sources of correlation which correspond to indirect rather than direct causation:  we speak then of spurious-correlation, correlation which can be explained by confounding factors, by common causes. 

There are serious reasons to doubt the uniformity of incidents over shifts. There may occur periodical differences. The population of a hospital ward may vary over the seasons. The patients may differ in character and severity of illness due to seasonal influences. There are differences between day and night shifts and between weekend shifts and shifts on weekdays. An extensive study of Dutch Intensive Care Units admissions shows a marked increase in deaths when the admission falls outside ``office hours''\cite{Kuijsten:2010}.  Recall that there have to be nurses on duty throughout the night and throughout the weekends, while the medical specialists tend to have ``normal working hours''. Finally there is the periodical cycle of the circadian rhythm, influencing the condition of the patients and the attention of the medical staff \cite{Kuhn:2000}. 

Notice that circadian variation in e.g. mortality and the resulting variation of incident rate between different shifts over the day interacts with the variation in the number of nurses on a shift, with more personnel on the day shifts. This can result in a higher number of nurses with an incident on their shift if the incident rate is higher during day time shifts and conversely, a lower number in the opposite case.

There may be other, non-periodical variations that affect the uniformity of incidents. In the case of the Juliana Children's hospital there has been a rather sensitive matter of policy: whether very ill children, who are not going to live for very long, should die at home or in the hospital wards. We understand that this policy did change at least once at the JCH in the period of interest. Presumably a change in policy concerning where the hospital wants children to die, will have impact on the rate of incidents. Further, incidents may be clustered, since one patient can give rise to several incidents. 

On the other hand the way nurses are assigned to shifts is certainly not uniform and `random'. Nurses take shifts in patterns, for example several night shift on a row, alternated by rows of evening or day shifts. Nurses are assigned to shifts according to skills, qualification and other characteristics. Maybe some nurses take relatively more weekend shifts than others, because of personal circumstances. 

Although both the assignment of nurses to shifts and the assignment of incidents to shifts are not uniform processes, one could hope that there might be some `mixing' condition that makes the ultimate result indistinguishable from the postulated independence and uniformity. Certainly one may hope, but this magical mechanism should at least be made plausible.

Taken together, even if we consider both the shifts of a given nurse as a random process, and the incidents on a ward as a random process, and even if we consider the two processes as stochastically independent of one another, the assumption of constant intensities of either is a guess, not based on any evidence or argument. There may be patterns in the risk of incidents and there are certainly patterns in the shifts of nurses. These patterns may be correlated, through the process by which shifts are shared over the different nurses according to their different personal situations, their different wishes for particular kinds of shifts, their different qualifications, and the changing situation on the ward.

\subsection*{How stable are the hypergeometric probabilities under small changes in the data?}
\label{subsection:robustness}
Consider the data  of the ward at JCH. 
These numbers and their interpretation are at the root of what turned out to be one of the gravest miscarriages of justice of the Dutch Juridical system. 
Under the assumption of the hypergeometric distribution the probability of this configuration is very small, less than one in nine million. The configuration is in some respects extreme: eight out of eight incidents occur in the shift of one nurse. However the data are in another respect also conspicuous: no incidents occur in the 887 shifts where this nurse was not present (see Table 1). 
The data collection had been far from flawless, with no formal definition of incident, no or incomplete documentation, and rested at least in part on recollection of witnesses who were aware of which facts were looked for.
 
Assuming the possibility of tiny flaws in the process of data acquisition, it is legitimate to investigate the effect of $1, 2,\dots, 8$ incidents that could have been forgotten, or overlooked. This amounts to allowing a maximal error of less than one percent. The results are quite remarkable; in table \ref{ref:table:stable} we give the probabilities..

\end{multicols}
\begin{table}[ht]
\caption{The effect of perturbations on the probabilities}
\begin{center} \footnotesize
\begin{tabular}{|l|c|c|c|c|c|}
\hline
Shifts with incidents&&&&& \\
outside Lucia's (postulated)&0&1&2&3&4 \\
\hline
Probability& 1/9043864& 1/1137586& 1/257538& 1/79497& 1/29989 \\
\hline
\hline
Shifts (continued)&5&6&7&8& \\
\hline
Probability& 1/13051 & 1/6329 & 1/3341& 1/1889& \\
\hline

\end{tabular}

\end{center}
\label{ref:table:stable}
\end{table}

\begin{multicols}{3}
The very small numbers vanish easily. Six or more incidents not remembered, not reported, or just defined away make the difference between astronomically small on the one hand and very unusual on the other. This shows that the probabilities are very sensitive to small errors in the data. 

A judgement on data quality is not only the concern of a statistician. Judges are used to inconsistent and incomplete data (statements), psychologists are very well aware of the possible fallacies of memory. Both groups have their own professional standards of how to deal with these phenomena. A statistician, however, should point out what the effects of these phenomena can be on the outcome of his models.

If this model is used to corroborate evidence this sensitivity should be made explicit, just as adverse workings of a medicine are mentioned explicitly for the users.

\wegermee{
\section*{Concluding remarks on the effects of heterogeneity}
\label{subsection:conclusion}
In the body of this paper we have shown the dramatic effect a modest amount of heterogeneity can have on tail probabilities, the probability that one nurse would experience a strikingly large amount of incidents. 
The data heterogeneity may stem from many different origins, be it variation  in the characteristics of nurses, variation in the circumstances in time, or correlation between these. Room for subjective bias in data acquisition is, among others, a form of variation over the nurses. One of the outcomes of this section is, rather surprisingly, that a simple discrete probability model like that of the hypergeometric distribution is based on assumptions that have to be verified, restricting its applicability in general.

Given the effects of heterogeneity on the outcomes, an analysis of possible sources of heterogeneity and their possible outcomes should accompany statistical reports, especially when there is so much at stake. 

We conjecture that a more open minded and careful analysis of the data on the early stages of what was to become the case ``Lucia de Berk" would have dampened the initial panic. The initial analyses which produced the astronomical small
probabilities have been decisive in creating a very narrow mindset and tipped off the disastrous course of events.
}

\section*{Concluding remarks}
\label{subsection:conclusion}
In the body of this paper we have shown the considerable effect that a modest amount of heterogeneity can have on tail probabilities. The broader impact of allowing heterogeneity in the analysis of (medical) research has interesting consequences outside the case of Lucia de Berk. 
What remains is a very short description of how the case ended in acquittal. Lucia was arrested in december 2001. As indicated in the introduction, the court (of appeal) stated that it did not include statistical considerations as basis for its verdict. This may be true for formal statistical considerations, but the essential step in the construction of the guilty verdict was that only one or two cases of murder had to be proven convincingly, the rest of the murders could be considered proven based on the "very improbable" occurrence of incidents during the shifts of Lucia. In this way statistical considerations were crucial, but the verdict was immunized against formal statistics. In this way Lucia was convicted in 2004 for seven murders and three attempts of murder. 
What followed was a long legal struggle where the emphasis was on the validity of the medical arguments and increasingly intricate juridical matters. 
The Lucia case was fiercely debated in public and the statistical notions remained here an important issue. Statisticians, now banned from the courtrooms, continued to play a role, for example by mobilizing the scientific community. Gradually the notion emerged that a gross miscarriage of justice had taken place. A complicating factor remained that, since the juridical path had been followed till the end, a new ``fact'', a so-called {\it novum} had to be found.
In 2008, Lucia was allowed to wait for the end of the legal proceedings outside prison, and two years later she was finally acquitted of all murder accusations.

\end{multicols}

\clearpage

\bibliographystyle{plainnat}
\bibliography{Lucia}
\end{document}